\documentclass[12pt,preprint]{aastex}

\shortauthors{Matthews et al.}
\shorttitle{HI Observations of $\delta$ Cep}

\begin{document}
\newcommand{\ang}{\rm \AA}
\newcommand{\msun}{M$_\odot$}
\newcommand{\lsun}{L$_\odot$}
\newcommand{\days}{$d$}
\newcommand{\degree}{$^\circ$}
\newcommand{\ud}{{\rm d}}
\newcommand{\as}[2]{$#1''\,\hspace{-1.7mm}.\hspace{.0mm}#2$}
\newcommand{\am}[2]{$#1'\,\hspace{-1.7mm}.\hspace{.0mm}#2$}
\newcommand{\ad}[2]{$#1^{\circ}\,\hspace{-1.7mm}.\hspace{.0mm}#2$}
\newcommand{\lsim}{~\rlap{$<$}{\lower 1.0ex\hbox{$\sim$}}}
\newcommand{\gsim}{~\rlap{$>$}{\lower 1.0ex\hbox{$\sim$}}}
\newcommand{\HA}{H$\alpha$}
\newcommand{\HII}{\mbox{H\,{\sc ii}}}
\newcommand{\kms}{\mbox{km s$^{-1}$}}
\newcommand{\HI}{\mbox{H\,{\sc i}}}
\newcommand{\KI}{\mbox{K\,{\sc i}}}
\newcommand{\HeI}{\mbox{He\,{\sc i}}}
\newcommand{\nan}{Nan\c{c}ay}
\newcommand{\galex}{{\it GALEX}}
\newcommand{\jks}{Jy~km~s$^{-1}$}
\newcommand{\DC}{$\delta$~Cephei}

\title{New Evidence for Mass-Loss from $\delta$~Cephei from \HI\ 21-cm
Line Observations}

\author{L. D. Matthews\altaffilmark{1},  
M. Marengo\altaffilmark{2}, N. R. Evans\altaffilmark{3}, \& G. Bono\altaffilmark{4,5}}

\altaffiltext{1}{Massachusetts Institute of Technology 
Haystack Observatory, Off Route 40, Westford, MA
  01886 USA}
\altaffiltext{2}{Department of Physics \& Astronomy, 
Iowa State University, Ames, IA 50011 USA}
\altaffiltext{3}{Harvard-Smithsonian Center for Astrophysics, 60
  Garden Street, MS-42, Cambridge, MA 02138 USA}
\altaffiltext{4}{Department of Physics, Universit\`a di Roma Tor
  Vergata via Della Ricerca Scientifica 1, 00133 Roma, Italy}
\altaffiltext{5}{INAF, Rome Astronomical Observatory, via Frascati 33,
00040 Monte Porzio Catone, Italy}

\begin{abstract}
Recently published {\it Spitzer Space Telescope} observations of the classical
Cepheid archetype $\delta$~Cephei revealed an extended
dusty nebula surrounding this star and its hot companion 
HD~213307. At far infrared
wavelengths, the emission
resembles a bow shock aligned with the direction of space motion of
the star, 
indicating that $\delta$~Cephei is undergoing mass-loss through a
stellar wind.  Here we report \HI\ 21-cm line observations with 
the Very Large Array (VLA) to search for 
neutral atomic hydrogen associated with this wind. 
Our VLA data reveal a spatially extended \HI\ nebula ($\sim13'$ or
1~pc across) surrounding the position of \DC. The nebula has a
head-tail morphology, consistent with circumstellar ejecta shaped by
the interaction between a stellar wind and the interstellar medium (ISM).
We directly measure a mass of circumstellar atomic
hydrogen $M_{\rm HI}\approx0.07~M_{\odot}$, although the total \HI\ mass may be
larger, depending on the fraction of circumstellar
material that is
hidden by Galactic contamination within our band or that is
present on angular scales too large to be detected by the VLA. 
It appears that the bulk of
the circumstellar 
gas has originated directly from the star, 
although it may be augmented by material swept
from the surrounding ISM.
The \HI\ data are consistent with a stellar wind with an outflow
velocity $V_{\rm o}=35.6\pm1.2$~\kms\ and
a mass-loss  rate of 
${\dot M}\approx(1.0\pm0.8)\times 10^{-6}~M_{\odot}$ yr$^{-1}$. 
We have computed theoretical 
evolutionary tracks that include mass loss across the
instability strip and show that
a mass-loss rate of this magnitude, sustained over
the preceding Cepheid lifetime of \DC, could be sufficient to resolve
a significant fraction of the
discrepancy between the 
pulsation and evolutionary masses for this star.

\end{abstract}


\section{Introduction\protect\label{introduction}}
Owing to the tight coupling between the pulsation properties of
Cepheid variables  and  
their fundamental stellar parameters,  these stars have
long provided crucial tests for stellar evolution
models (Gautschy \& Saio 1996) 
and play a key role in the calibration of the extragalactic
distance scale (Feast \& Walker 1987; Freedman et al. 2001).  
Nonetheless,
for more than 40 years, a puzzle has persisted concerning the
lingering discrepancy between stellar masses derived for Cepheids
using different methods (e.g., Cox 1980). In spite of improved radiative 
opacity calculations
(Iglesias et al. 1990; 
Seaton et al. 1994), advances in evolutionary modeling (Bono
et al. 2002), and the consideration of metallicity effects (Keller
\& Wood 2006), masses derived from
stellar evolutionary models are found to be 
systematically $\sim$10-15\% higher than
those derived from stellar pulsation models (Caputo et al. 2005)
or orbital dynamics, when
available (Evans 2009). 

As described by Neilson et al. (2011), one of the most promising
solutions to the so-called ``Cepheid mass discrepancy'' 
is likely to be convective
core overshooting (Chiosi et al. 1992) coupled with {\em mass-loss} during the
Cepheid phase of evolution (see also Bono et al. 2006).
Mass-loss from Cepheids has long been postulated, not only as a means of
resolving the Cepheid mass discrepancy, but as a natural consequence
of stellar evolution and pulsation 
models (e.g., Iben 1974; Willson \& Bowen 1984; 
Neilson \& Lester 2008). 
The observational confirmation of mass loss from Cepheids would impact not only
our understanding of
Cepheid evolution, but would also have
implications for the use of Cepheids as
distance indicators. For example, 
mass-loss can non-negligibly affect the structure and scatter in the infrared
period-luminosity relation (Neilson et al. 2009), and the presence of
circumstellar material can impact Cepheid distance determinations made using
the interferometric Baade-Wesselink method (M\'erand et al. 2007).

A number of previous authors have attempted to identify direct
evidence of past or ongoing mass-loss from Cepheids using
observations ranging from ultraviolet to radio wavelengths (e.g.,
Deasy \& Butler 1986; 
Deasy 1988; Welch \& Duric 1988; Bohm-Vitense \& Love 1994; M\'erand
et al. 2007; Kervella et al. 2006, 2009; Neilson et
al. 2009). 
However, the results have been
inconclusive, as the mass-loss rates (or upper limits) 
derived from these studies (${\dot M}\approx 
10^{-12}$ to $10^{-5}~M_{\odot}$~yr$^{-1}$)
span many orders of magnitude and are typically rather uncertain owing
to their reliance on trace atomic and molecular species (which may
not be optimal tracers of past and ongoing mass-loss), their inability
to sample very extended spatial scales, and finally, the
strong dependencies of derived mass-loss rates 
on  a variety of underlying assumptions (e.g., wind structure,
gas-to-dust ratios,
ionization fractions).

With the goal of obtaining new empirical constraints on Cepheid mass-loss,
Marengo et al. (2010a) recently used the {\it Spitzer
Space Telescope} to
survey a sample of 29 nearby Cepheids. 
While IR excesses were not directly detected, 
ruling out the presence of a large amount of warm ($\sim$500~K) dust in close
proximity to the stars, the {\it Spitzer} images
revealed extended
emission around several targets (Marengo et al. 2010a; Barmby et
al. 2011).
Perhaps the most intriguing was the discovery of
a prominent nebula around the archetype Cepheid variable,
\DC\ (Marengo et al. 2010b).  

Some basic stellar properties of \DC\ are summarized in Table~1. \DC\
is part of a wide binary (Benedict et al. 2002), 
with a hot 
companion HD~213307 ($T_{\rm eff}=$8800~K; Cenarro et al. 2007) 
separated by a projected distance of 
$40''$. The IR emission surrounding \DC\ (and
HD~213307)
is visible in multiple {\it Spitzer} bands
and shows a roughly parabolic structure with an extent of
$\sim5'$ ($\approx2\times10^{4}$~AU).  
The symmetry axis of the
parabola is aligned with the direction of the star's motion
through the interstellar medium (ISM; Marengo et al. 2010b), where 
$V_{\rm space}\approx 10.3$~\kms\ and P.A.=\ad{58}{3}.\footnote{For
  the present work, we have recomputed 
the space motion vector for \DC\ using the updated
solar constants from Sch\"onrich et al. 2010.} The IR emission
therefore appears to trace
a bow shock structure, as arises when a moving,
mass-losing star interacts with the local ISM
(e.g., Wilkin 1996).
Bow shocks have been seen previously in FIR
images of a number of mass-losing stars, including the 
supergiants $\alpha$~Cam 
(Van Buren \& McCray 1988) and $\alpha$~Ori  
(Noriega-Crespo et al. 1997; Ueta et
al. 2008) and the asymptotic giant
branch (AGB) stars R~Hya (Ueta et al. 2006; Wareing et al. 2006), Mira
(Ueta 2008), 
R~Cas (Ueta et al. 2010), IRC+10216 (Ladjal et al. 2010), TX~Psc, and
X~Her (Jorissen et al. 2011). 
The emission mechanism responsible for the FIR emission
from these circumstellar bow shocks is
uncertain, but is likely to be mainly thermal emission from dust,
with possible contributions from low-excitation atomic emission lines
(Ueta et al. 2008; Marengo et al. 2010b). 

While the
companion of \DC, HD~213307, cannot strictly be excluded as the source
of the IR emitting material seen by {\it Spitzer}, 
this appears quite unlikely based on
the B7-B8~III-V spectral type of this star (see Marengo et al. 2010b for
discussion). 
Thus the {\it Spitzer} observations of Marengo et al. (2010b) provide
strong evidence that
\DC\ is undergoing mass loss through a stellar wind. To 
further characterize the nature of this wind and
its interaction with the ISM, 
we have now used the Very Large Array (VLA) to search for \HI\
21-cm line emission in the circumstellar environment of \DC. 

\HI\ observations have been  used previously to 
trace extended circumstellar emission surrounding a number of
mass-losing AGB stars (e.g., G\'erard \&
Le~Bertre 2006; Matthews \& Reid 2007; Libert et al. 2008; 
Matthews et al. 2008,
2011). Frequently, the \HI\ emission associated with AGB stars 
is highly extended (up to $\sim$1~pc) and shows 
signatures of interaction with the ISM in the form of trailing \HI\ 
wakes, velocity gradients caused by ram pressure effects, and/or
density enhancements that demarcate
the interstellar-circumstellar interaction zone. Furthermore, several of the
evolved stars detected in \HI\ are among those 
with FIR (and/or FUV) emitting bow shocks, including Mira,
R~Cas, and IRC+10216, underscoring the complementarity of these
tracers for probing the chemistry and kinematics of
stellar outflows and their interaction with their environments.
This paper represents the first extension of this approach to the
study of Cepheids.

\begin{deluxetable}{lcc}
\tabletypesize{\scriptsize}
\tablewidth{0pc}
\tablenum{1}
\tablecaption{Coordinates and Stellar Properties of $\delta$~Cephei}
\tablehead{\colhead{Parameter} & \colhead{Value} & \colhead{Ref.}}
\startdata

$\alpha$ (J2000.0) & 22 29 10.2 & 1\\

$\delta$ (J2000.0) & +58 24 54.7 & 1 \\

$l$ & \ad{105}{19} & 1 \\

$b$ & \ad{+0}{53} & 1\\

Distance (pc) & 273$\pm$0.011 & 2\\

Spectral Type & F5Ib-G1Ib & 3\\

Pulsation period (days) & 5.366341 & 3\\

Mean $T_{\rm eff}$ (K) & 5910 & 4\\

Mass (pulsation)$^{b}$ ($M_{\odot}$) & 4.5$\pm$0.3 & 5\\

Mass (evolutionary)$^{c}$ ($M_{\odot}$) & 5.7$\pm$0.5 & 5\\

Mean $M_{V}$ & $-$3.47$\pm$0.10 & 2 \\

Mean Luminosity ($L_{\odot}$)& $\sim$2000 & ...\\

Mean radius ($R_{\odot}$) & 44.5$^{a}$ & 6\\

$V_{\rm LSR}^{d}$ & $-4.7$~\kms\ & 7 \\

\enddata

\tablenotetext{a}{Based on the mean, limb darkened angular diameter 
$\phi_{\rm LD}$=1.520~mas and the distance adopted here.}

\tablenotetext{b}{Based on $V-K$ colors.}

\tablenotetext{c}{Assuming a canonical model with no overshoot; see
  \S~\ref{resolution}).} 

\tablenotetext{d}{Derived from the heliocentric
radial velocity $V_{\rm h}=-16.8$~\kms.}

\tablecomments{Units of right ascension are hours, minutes, and
seconds, and units of declination are degrees, arcminutes, and
arcseconds. All quantities have been scaled to the distance adopted in
this paper.}

\tablerefs{(1) SIMBAD database; (2) Benedict et al. 2002; 
(3) Samus et al. 2011;  (4)
Andrievsky et al. 2005; (5) Caputo et
al. 2005; (6) Armstrong et al. 2001; (7) Wilson 1953 }

\end{deluxetable}

\section{VLA Observations\protect\label{VLAobs}}
\HI\ 21-cm line observations of \DC\ were obtained using the VLA of
the National Radio Astronomy Observatory (NRAO)\footnote{The National
Radio Astronomy Observatory is operated by Associated Universities,
Inc., under cooperative agreement with the National Science
Foundation.} during four observing sessions in 2009 (see Table~2).
All data were obtained using the compact (D)
configuration (0.035-1.0~km baselines), providing
sensitivity to emission on scales of up to $\sim$15$'$. The
primary beam of the VLA at the observing
frequency of 1420.5~MHz is $\sim 31'$. 

The VLA correlator was used in 1A
mode with a 1.56~MHz bandpass, yielding 512 spectral
channels with 3.05~kHz ($\sim$0.64~\kms) spacing in a single (right
circular) polarization. The band was centered
at a velocity of $-7.5$~\kms\ relative to the local standard of rest
(LSR).
In total, $\sim$15.1 hours of integration were obtained on
\DC. Approximately 20\% of the observed visibilities were
flagged because of radio frequency interference (RFI) or 
hardware problems. 
Observations of \DC\ were interspersed with observations of a phase
calibrator, 2203+626, approximately every 20
minutes. 3C48 (0137+331) was used as a flux calibrator, and an
additional strong point source (2253+161) was observed as a bandpass 
calibrator (Table~2). To 
insure that the absolute flux scale and bandpass calibration were not
corrupted by Galactic emission in the band, the
flux and bandpass calibrators were each observed twice,
with frequency shifts of $+$1.1~MHz and $-1.7$~MHz, respectively,
relative to the band center used
for the observations of \DC\ and 2203+626. 2203+626 was
also observed once at each of these offset frequencies to
permit more accurate bootstrapping of the absolute 
flux scale  to the \DC\ data, although in some cases, one of the offset
scans had to be discarded (see Table 2). 
We estimate that the resulting absolute flux scale has an
uncertainty of $\sim$10\%.

The data were calibrated and imaged
using the Astronomical Image Processing System
(AIPS). 
At the time of our
observations, the VLA contained 22 operational antennas with L-band
receivers, 20 of which had
been retrofitted as part of the Expanded Very Large Array (EVLA)
upgrade. Data obtained during this EVLA transition period require special
care during calibration.\footnote{See 
http://www.vla.nrao.edu/astro/guides/evlareturn/.}
After applying the latest available corrections to the 
antenna positions and performing an initial excision of corrupted data, 
we  computed and applied a bandpass calibration to our
spectral line data to remove closure errors on 
VLA-EVLA baselines. The bandpass was normalized using channels
163-448, thus excluding the portion of the band affected by aliasing.
We next computed
a frequency-averaged (channel~0) data set for use in calibrating
the frequency-independent complex gains, again using channels 163-448. 
Following gain calibration, we 
applied time-dependent frequency
shifts to the data to compensate for changes caused by
the Earth's motion during the course of the observations. At this
stage, we also applied hanning smoothing in velocity and
discarded every other channel, yielding a 256 channel data set with a
velocity resolution of $\sim$1.3~\kms.

Prior to imaging the line data, the $u$-$v$ data were
continuum-subtracted using a zeroth order fit to the 
real and imaginary
components of the visibilities. Channels 20-95 
of the hanning-smoothed data set (corresponding to 
an LSR velocity range of 35.6 to
132.2~\kms)
were determined to be line-free and were used for these
fits.  Although the spectral shape of the
aliased portion of the
continuum was 
better approximated by a higher order polynomial, the weakness of the
continuum in the line data coupled with the lack of line-free channels
on the high-frequency end of the band did not provide adequate
constraints for a higher order fit.  

We imaged the \HI\ line data using the standard AIPS CLEAN deconvolution 
algorithm and produced data cubes using various 
weighting  schemes. Some characteristics of these cubes are summarized in 
Table~3. 
Additionally,
we produced an image of the 21-cm continuum
emission in the \DC\ field using the line-free portion
of the
band. The peak continuum flux density 
within the primary beam was
$\sim$0.5~Jy (after correction for beam attenuation). 
We compared the measured flux densities of several
sources in the field with those from the NRAO VLA Sky Survey (Condon
et al. 1998) and found agreement to within $\lsim$7\%.
We did not detect any continuum emission
at or near the position of \DC\ and place a 3$\sigma$ upper limit on 
the 21-cm
continuum flux density at the stellar position of $F_{\rm
  cont}<$1.0~mJy.
This is consistent with the results of Welch \& Duric (1988), who
found no evidence of significant mass-loss from \DC\ in the form of an
ionized wind based on radio continuum observations at 5~GHz.

%
\begin{deluxetable}{lccll}
\tabletypesize{\scriptsize}
\tablewidth{0pc}
\tablenum{2}
\tablecaption{VLA Calibration Sources}
\tablehead{
\colhead{Source} & \colhead{$\alpha$(J2000.0)} &
\colhead{$\delta$(J2000.0)} & \colhead{Flux Density (Jy)} & \colhead{Date}
}

\startdata
3C48$^{\rm a}$      & 01 37 41.29 & +33 09 35.13 & 15.88$^{*}$   &
All\\
2253+161$^{\rm b}$& 22 53 57.74 &+16 08 53.56  &
13.74$\pm$0.15$^{\dagger}$ & 2009-Oct-28\\
...       & ...           & ...           & 13.46$\pm$0.28$^{**}$  &
2009-Nov-15\&16\\
...       &...            & ...           & 13.27$\pm$0.23$^{**}$ &
2009-Nov-22\&23\\
...       & ...           & ...           & 13.36$\pm$0.22$^{\dagger}$ &
2009-Nov-27\&28\\
2203+626$^{\rm c}$  & 22 03 20.96 & +62 40 34.27 &2.75$\pm$0.02$^{**}$ &
2009-Oct-28\\
...       & ...           & ...           & 2.72$\pm$0.01$^{**}$ 
& 2009-Nov-15\&16\\
...       & ...           &...            & 2.73$\pm$0.01$^{**}$ &
2009-Nov-22\&23\\
...       & ...           & ...           &
2.75$\pm$0.01$^{\dagger\dagger}$ 
& 2009-Nov-27\&28\\
\enddata

\tablecomments{Units of right ascension are hours, minutes, and
seconds, and units of declination are degrees, arcminutes, and
arcseconds. }
\tablenotetext{*}{Adopted flux density at 1420.5~MHz,
computed according to the VLA
Calibration Manual (Perley \& Taylor 2003).}
\tablenotetext{\dagger}{Computed flux density at 1419.2~MHz; data at
  1422.1~MHz were corrupted.}
\tablenotetext{**}{Mean computed flux density from 
observations at 1419.2 and 1422.1~MHz; see \S~\ref{VLAobs}.}
\tablenotetext{\dagger\dagger}{Computed flux density at 1422.1~MHz;
  data at 1419.2~MHz were corrupted.}

\tablenotetext{a}{Primary flux calibrator}
\tablenotetext{b}{Bandpass calibrator}
\tablenotetext{c}{Phase calibrator}

\end{deluxetable}

%
\begin{deluxetable}{lccccc}
\tabletypesize{\footnotesize}
\tablewidth{0pc}
\tablenum{3}
\tablecaption{Deconvolved Image Characteristics}
\tablehead{
\colhead{Image} & \colhead{{$\cal R$}} & \colhead{Taper} & 
\colhead{$\theta_{\rm FWHM}$} & \colhead{PA} & \colhead{rms} \\ 
\colhead{Descriptor} & \colhead{} & \colhead{(k$\lambda$,k$\lambda$)} & 
\colhead{(arcsec)} & \colhead{(degrees)} & \colhead{(mJy
beam$^{-1}$)} \\
\colhead{(1)} & \colhead{(2)} & \colhead{(3)} &
\colhead{(4)} & \colhead{(5)} & \colhead{(6)}  }

\startdata

Robust +1 & +1 & ... & $49''\times45''$ & \ad{20}{3} & 1.6\\

Natural & +5 & ...& $54''\times49''$ & \ad{21}{7} & 1.5\\

Tapered & +5 & 2,2 & $97''\times89''$ & \ad{2}{5}  & 1.8 \\

Continuum & $-$1 & ... & $39''\times36''$ & \ad{0}{3} & 0.39\\


\enddata

\tablecomments{
Explanation of columns: (1) image or data cube designation 
used in the text; (2) AIPS robustness
parameter used in image deconvolution; 
(3) Gaussian taper applied in the $u$ and
$v$ directions, expressed as
distance to 30\% point of Gaussian in units of kilolambda;
(4) dimensions of
synthesized beam; (5) position angle of synthesized beam (measured
east from north); (6) mean rms
noise per channel (1~$\sigma$) in the unaliased portion of the band 
(line data) or in frequency-averaged
data (continuum).}

\end{deluxetable}

\section{Results\protect\label{results}}
\begin{figure}
\vspace{-1.5cm}
\centering
\scalebox{0.9}{\rotatebox{0}{\includegraphics{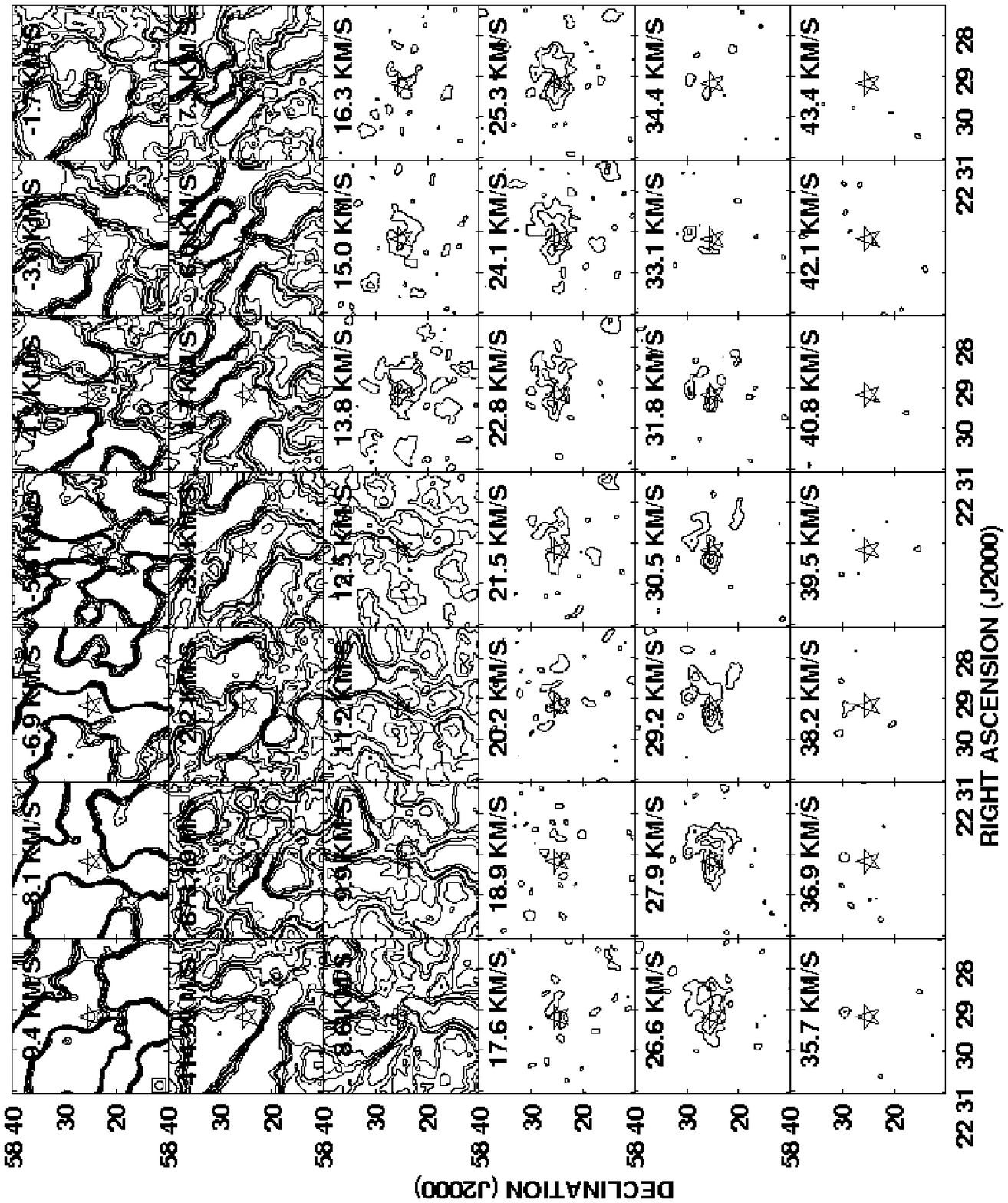}}}
\caption{Selected \HI\ channel maps 
from the Tapered VLA data (see Table~3). 
A star symbol indicates the position of \DC.
Contour levels are 
($-12$,$-6$,$-$3,3,6,12)$\times$1.83~mJy beam$^{-1}$. The lowest
contour is $\sim3\sigma$. The size of
the synthesized beam is indicated in the lower corner of the upper
left panel. Channels with $V_{\rm LSR}\lsim$13~\kms\
are dominated by large-scale Galactic emission that is
poorly spatially sampled by the VLA, leading to a strongly mottled
appearance. For this reason, only a few
representative channels blueward of the stellar systemic velocity ($V_{\rm
LSR}=-4.7$~\kms) are shown.
  }
\label{fig:cmaps}
\end{figure}

\begin{figure}
\vspace{-1.5cm}
\centering
\scalebox{0.9}{\rotatebox{0}{\includegraphics{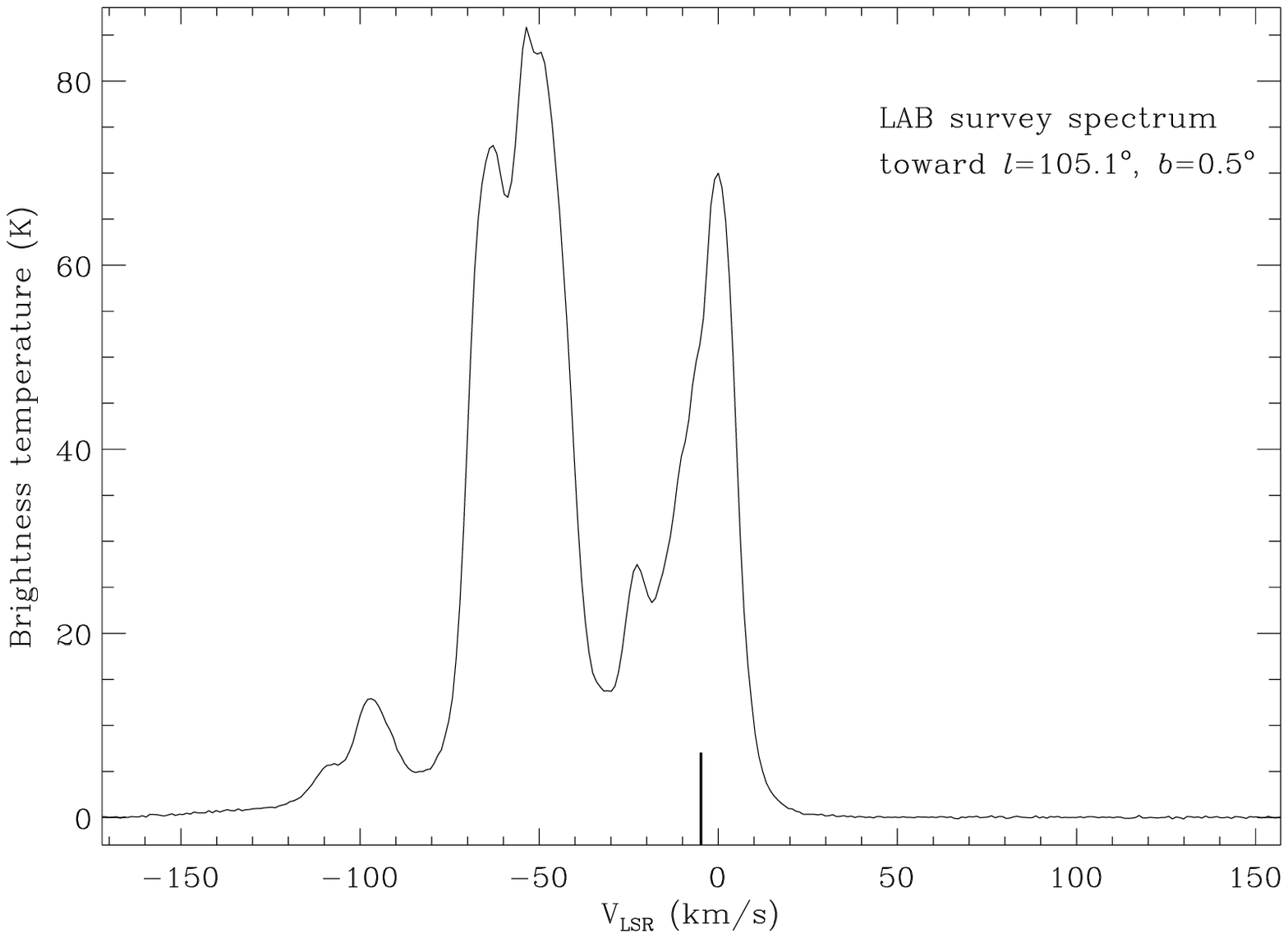}}}
\caption{Single-dish \HI\ spectrum toward $l=$\ad{105}{1},
$b=$\ad{0}{5}, illustrating the nature of the Galactic  
emission along the line-of-sight to \DC. The data were taken from
Kalberla et al. 2005.  
The velocity range shown corresponds to our VLA observing
band. The vertical bar indicates the systemic velocity of \DC.
Note that the rms noise level in 
this spectrum  ($\sim0.09$~K or $\approx$0.57~Jy) 
is too high to permit detection of circumstellar emission even in
the uncontaminated portion of the band.  }
\label{fig:surveyspec}
\end{figure}

Figure~\ref{fig:cmaps} shows a series of \HI\ channel maps extracted
from our Tapered VLA data cube (see Table~3).  Only a
few representative channels are shown blueward of the stellar systemic velocity
($V_{\rm LSR}=-4.7$~\kms), as this portion of our band was strongly
contaminated by large-scale Galactic emission
along the line-of-sight. This emission is poorly spatially sampled by the VLA,
resulting in patterns of positive and negative mottling
across the channel images.  To better illustrate the character of the
Galactic emission,
Figure~\ref{fig:surveyspec} presents a single-dish \HI\ survey
spectrum from Kalberla et al. (2005), which 
shows \HI\ emission with a brightness
temperature as high as $\sim$85~K 
toward the direction of \DC. Both the strength and spatial complexity of this emission
hampered our search for  circumstellar
signals over the velocity range $-125\lsim V_{\rm LSR} \lsim
12.5$~\kms.

Fortunately, the spectrum in 
Figure~\ref{fig:surveyspec} reveals that the Galactic \HI\ emission
drops precipitously for $V_{\rm LSR}\gsim$13~\kms, providing a clean
band over which to search for circumstellar 
emission. Because the outflow velocities from
Cepheids are expected to be at least a few tens of \kms\ (e.g.,
Deutsch 1960; Reimers 1977;
Holzer \& MacGregor 1985), this
velocity window is of prime 
interest for searching for circumstellar emission associated with
\DC.

Consistent with the total power spectrum in Figure~\ref{fig:surveyspec},
we see in Figure~\ref{fig:cmaps} that the pattern in the VLA images
caused by large-scale Galactic emission diminishes near
$V_{\rm LSR}=13.8$~\kms. Furthermore, that same channel exhibits an
extended \HI\ emission region centered near the position of \DC. Looking toward
higher velocities, we find 
a series of 15 contiguous channels 
where spatially extended \HI\ emission is detected
coincident with the
position of \DC. Several marginally extended 
features of $\sim3\sigma$ significance are also seen clustering 
near the stellar position in each of the next five contiguous channels,
out to $V_{\rm LSR}=38.2$~\kms. 

\begin{figure}
\hspace{-0.8in}
\scalebox{0.4}{\rotatebox{-90}{\includegraphics{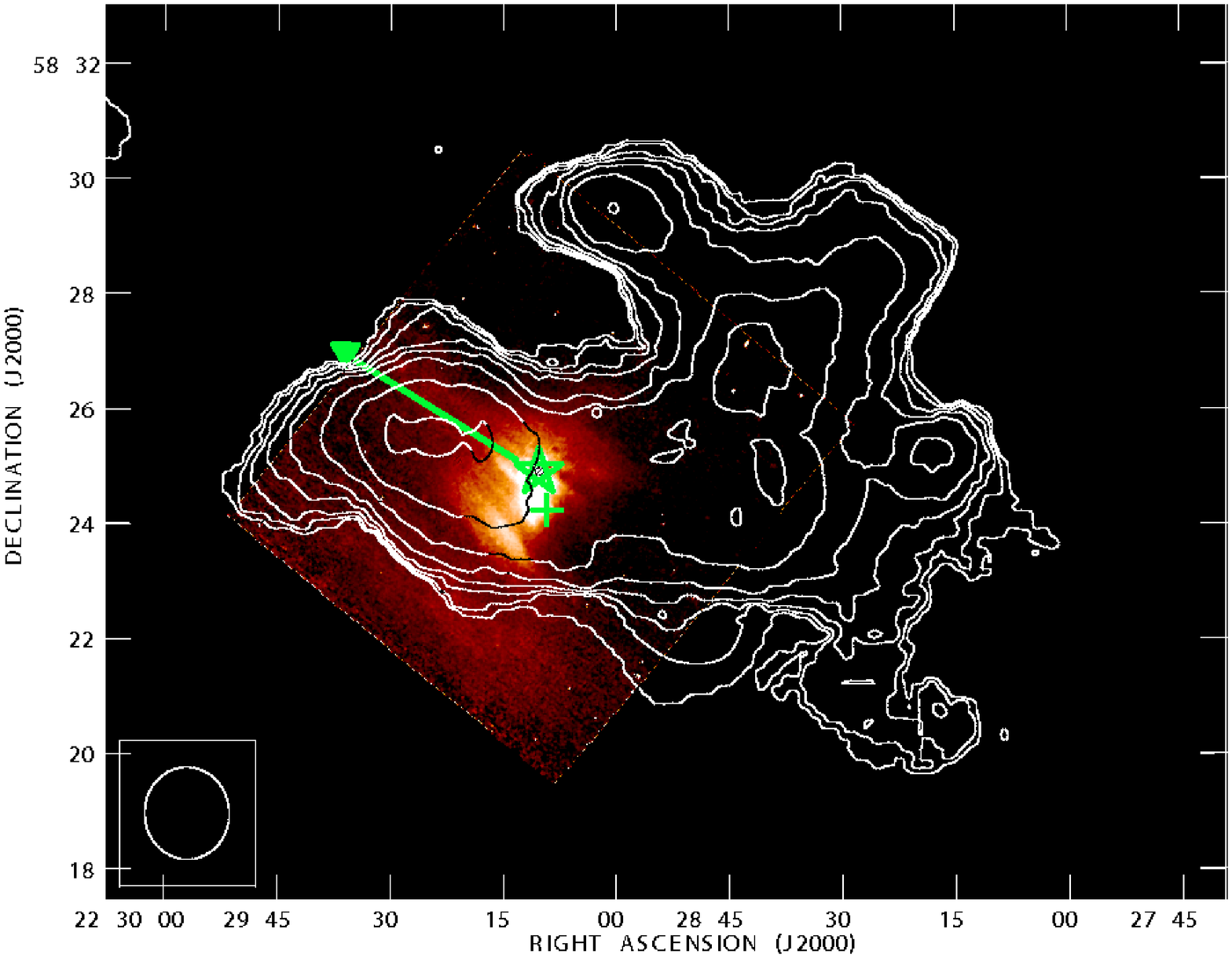}}}
\hspace{-0.7in}
\scalebox{0.4}{\rotatebox{-90}{\includegraphics{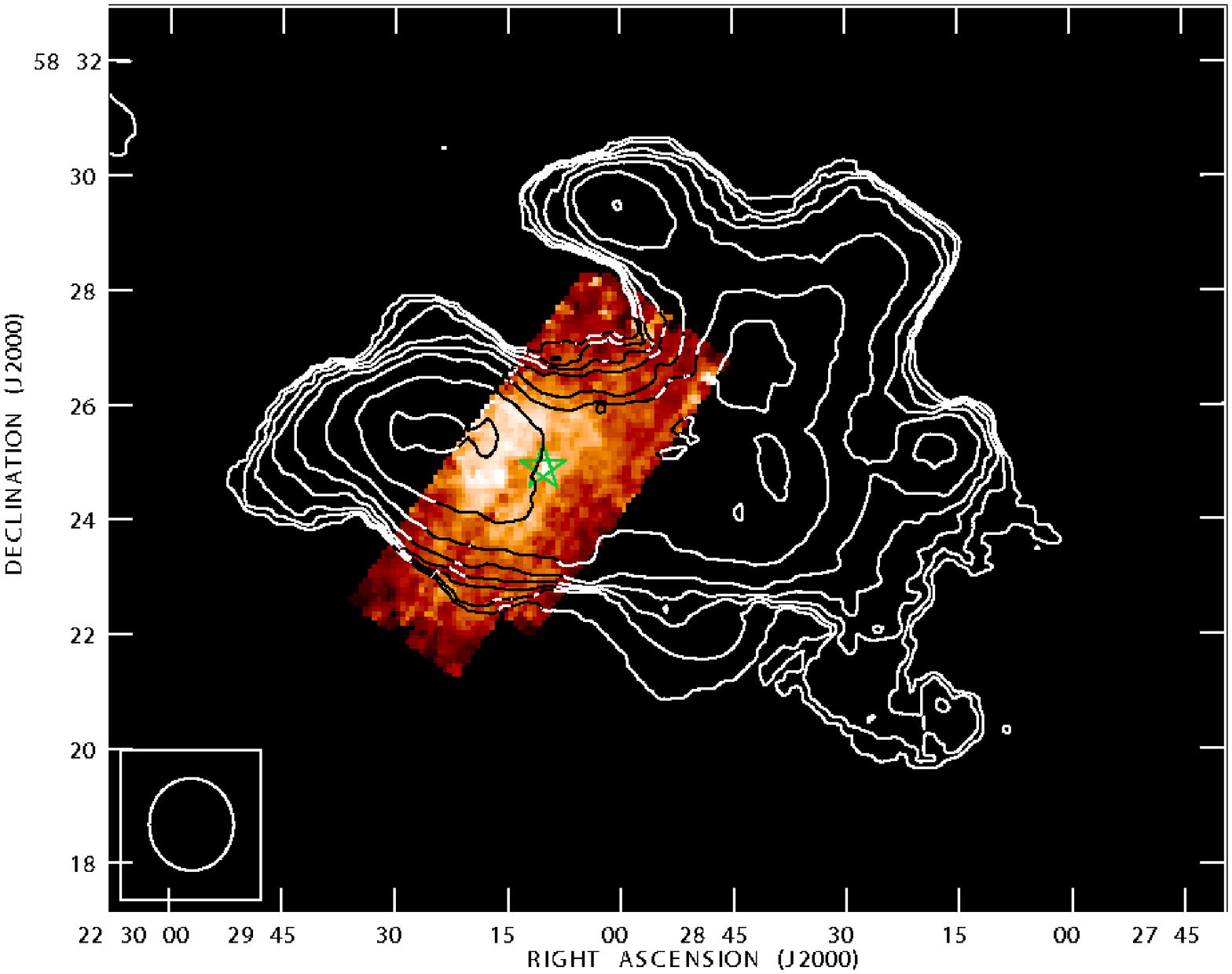}}}
\caption{\HI\ total intensity contours overlaid on {\it Spitzer} 
  24$\mu$m (left) and 70$\mu$m (right) images from Marengo et
  al. 2010b.  The \HI\ contour levels are
  (1,1.4,2...11.3)$\times$18~Jy~beam$^{-1}$ m s$^{-1}$. The positions
  of \DC\ and its companion HD~213307 are indicated by star and plus
  symbols, respectively. The
direction of space motion of \DC\ with respect to the
  ISM is indicated by a green arrow on the left panel. The insets in
  the lower left corners of each panel 
indicate the size of the VLA synthesized beam
  ($96''\times88''$). }
\label{fig:mom0}
\end{figure}

In Figure~\ref{fig:mom0} we present velocity-integrated \HI\ total
intensity contours, derived from data spanning the velocity
range $13.8\le V_{\rm LSR} \le 38.2$~\kms. These contours are 
overlaid on the {\it
  Spitzer} 24$\mu$m and 70$\mu$m maps, respectively, 
from Marengo et al. (2010b). 
Some caution is required in
interpreting the detailed morphology of the
velocity-integrated \HI\ distribution in Figure~\ref{fig:mom0}, 
as additional material associated with this nebula is likely to be 
present at lower velocities, where it cannot be disentangled
from the prominent Galactic emission. 
However, despite this caveat,
several characteristics of the integrated \HI\ emission seen in
Figure~\ref{fig:mom0} are 
strongly indicative of an association
with the circumstellar environment of \DC.  

One of the most intriguing aspects of the \HI\ emission distribution
in Figure~\ref{fig:mom0} is its ``head-tail'' morphology. This is
reminiscent of the structures of the extended circumstellar envelopes
of several of the 
AGB stars that have been imaged in \HI, FIR and/or
FUV light (Martin et al. 2007; 
Matthews et al. 2008, 2011; Ueta 2008; Sahai \& Chronopoulos
2010; Jorissen et al. 2011) 
and results when material shed by a stellar wind is 
swept by ram pressure into a trailing wake as the star transverses the
ISM (e.g., Wareing et al. 2007b). 
The direction of space motion of \DC, indicated by an arrow on
Figure~\ref{fig:mom0}, strongly supports an analogous interpretation 
for the \HI\
nebula surrounding this star. We see that the maximum angular extent
of the \HI\ nebula ($\sim13'$, corresponding to $\sim$1.0~pc at the distance of
\DC) is aligned with the space motion direction. 
Furthermore, the distribution of the \HI\ nearest to the
position of \DC\  ($r\lsim 2'$) is well correlated with
the distributions of both the 24$\mu$m and 70$\mu$m emissions detected by
{\it Spitzer}. 
The \HI\ emission thus appears to be associated with the same
dusty wind traced by the IR data.

The peak of the \HI\ column density in Figure~\ref{fig:mom0} ($N_{\rm
HI}\approx 2.0\times10^{19}$~cm$^{-3}$) does
not correspond to the position of 
\DC\ or its companion HD~213307 (indicated
on Figure~\ref{fig:mom0} by a star and cross symbol, respectively), but
rather lies $\sim112''$ to the northeast of \DC, near the apex of the shock
structure delineated by the IR emission. Such an offset of the peak
emission from the star is not
surprising, as this kind of leading density
enhancement is a hallmark of the
``snowplow'' effect that may occur when a wind-emitting star moves 
supersonically  through the ambient medium
(Isaacman 1979; Raga et al. 2008). 

To the west of \DC, i.e., downstream from the star's trajectory, 
our VLA image reveals a wake of material trailing the motion of the 
star. In contrast to the gaseous wakes associated with 
stars such as Mira (Matthews et al. 2008) and X~Her (Matthews
et al. 2011), the material downstream from \DC\ is not well collimated
and more closely resembles the wide-angle ``vortical tail''
associated with the carbon star IRC+10216 (Sahai \& Chronopolous
2010).  Numerical simulations of the interactions between stellar
winds and streaming environments frequently predict broad, turbulent
wakes, qualitatively similar to what we see in Figure~\ref{fig:mom0}
(e.g., Wareing et al. 2007a), although
the degree of collimation and other properties of circumstellar wakes
are highly sensitive to the stellar velocity, properties of the stellar
wind, and the density of the ambient medium (Comer\'on \& Kaper
1998), such that fully unraveling the interplay between these factors for
a particular source
generally requires a specifically
tailored model. 

To gauge the total mass of \HI\ giving rise to the emission seen in
Figures~\ref{fig:cmaps} and \ref{fig:mom0}
over the velocity range $V_{\rm LSR}$=15.0 to 38.2~\kms, we have summed
the emission surrounding the stellar position (after correction for primary
beam attenuation) in irregularly shaped blotches
defined by 3$\sigma$ brightness contours. This approach
yielded an integrated \HI\ flux density of $\int S_{v}dv=4.1\pm
0.2$~Jy~\kms. Using the relation $M_{\rm
  HI}=2.36\times10^{-7}d^{2}\int S_{v}dv~M_{\odot}$ 
where $d$ is the stellar distance in pc (e.g., Roberts 1975), this
translates to an \HI\ mass of $M_{\rm HI}\approx$0.07$M_{\odot}$.  
While this approach for measuring the total \HI\
context minimizes the contribution
of background noise to the measurement, it may result in the exclusion
of weak emission or 
emission from regions not
contiguous with the primary emission region in each channel. For comparison,
we therefore summed the emission in each channel over a
fixed rectangular aperture (\am{16}{6}$\times$\am{13}{3}, centered
100$''$ west of \DC's position). We found $S_{v}=4.4\pm0.4$~Jy~\kms,
consistent with our first estimate. 

We interpret our measured integrated \HI\ flux density as a lower limit to the
amount of circumstellar material 
present around \DC\ for two reasons. First,
additional emission may lie undetected at 
velocities where our spectral images were contaminated by Galactic 
emission. For example, if 
we assume that the \HI\ line profile is
symmetric about the systemic velocity of \DC, this would imply that 
roughly two-thirds of the channels containing circumstellar signal
were excluded from our flux density measurement. 
Secondly, at our observing frequency, the VLA D configuration is
only sensitive to emission on angular scales $\lsim 15'$. Since
this is comparable to the angular extent of the observed \HI\ nebula
surrounding \DC, it is plausible that we have missed additional
extended emission. This problem could be exacerbated if the most
spatially extended material is decelerated as a result of its interaction with
the ISM (e.g., Matthews et al. 2008); this could in turn cause its
velocity to be
blueshifted into the portion of our band contaminated by Galactic
emission. Mapping the region in \HI\ with a single-dish telescope
would be of considerable interest as a means of searching for 
an additional large-scale emission component.

\section{Discussion\protect\label{discussion}}
\subsection{How Robust is the Association between the \HI\ Nebula and
  \DC?}
The Galactic ISM is well known to exhibit
structures on wide variety of scales, down to arcminutes and below 
(e.g., Verschuur 1974; Greisen \& Liszt 1986; 
Heiles \& Troland 2003; Begum et
al. 2010). We therefore have considered the possibility that the
\HI\ nebula that we have detected toward \DC\ could simply be a
chance superposition of an unrelated \HI\ cloud along the
line-of-sight. However, several lines of evidence indicate that this
is extremely unlikely.

As described above, the close correspondence between the \HI\ emission contours
surrounding \DC\ 
and the bow shock structure traced by FIR emission implies that the
two are linked. Since the formation of a bow shock requires a stellar wind,
a linkage between the \HI\ and FIR emission argues strongly against the former
arising from
a random interstellar cloud.

Based on a survey with the Arecibo $L$-band Feed Array, Begum et
al. (2010) identified a population of compact \HI\ clouds associated
with the Galactic plane (but distinct from the main Galactic disk
emission) whose column densities and angular sizes
overlap with those characterizing 
the \HI\ nebula that we have detected toward \DC. 
However, the mean characteristic linewidth of the clouds
catalogued by Begum et al. is $\Delta V$=4.2~\kms\ (FWHM), with values
generally lying in the range of $\sim$1-8~\kms. Only a handful of
the 96 clouds in their sample exhibit velocity widths comparable
to or larger than that of the integrated linewidth of 
the \DC\ nebula ($\Delta V\approx 12.6$~\kms\ based on a Gaussian fit). 
Furthermore, while Begum et al. have underscored the possibility that a
subset of their compact clouds may in fact be associated with
mass-loss from variable stars, the closest association that they
identified between a catalogued variable and a cloud from their
sample has a projected
separation of 10$'$. Since the \HI\ properties of the \DC\ nebula are
somewhat atypical of the general population of compact Galactic
clouds, we conclude that 
the probability of detecting such a cloud with 
a head-tail
morphology
at both the position and velocity expected for \DC, is
extraordinarily small.

\subsection{The Stellar Mass-Loss Rate\protect\label{rates}}
\subsubsection{New Estimates of ${\dot M}$ 
Based on the VLA Data\protect\label{vlamassloss}}
One of the most powerful applications of our VLA measurements is that they
allow us to obtain new, independent constraints
on the mass-loss rate from \DC.  
Two key advantages of \HI\ observations for
characterizing stellar mass-loss are first, that the \HI\ mass can be
derived directly from the observed emission 
(assuming it is optically thin) and secondly, that spectroscopic
imaging of \HI\ is sensitive to
material much farther from the star than most other observational
techniques for probing stellar mass loss,
thereby tracing a larger fraction of the  total mass-loss history.

We consider two independent approaches for estimating the mass loss
rate of \DC\ from our VLA data.
Our first method is to adopt a simplified
model in which the star is assumed to be undergoing mass-loss through
an isothermal, spherically symmetric wind that is in free expansion
out to some radius $r=R_{0}$. While pulsation-driven
winds are inherently non-steady, we assume that the outflow can be
approximated as quasi-steady when averaged over the timescales sampled by
our observations.

In the case where the \HI\ emission is optically thin,
a theoretical line profile as a function of velocity, $V$, 
for this idealized wind can be written as:

\begin{equation}
 S(V) = \frac{2 f\sqrt{\pi {\rm ln}~2}~{\dot M}}
{4d\pi B V^{2}_{\rm o}(1.83\times10^{18})\mu m_{H}\left[1 -
    \left(\frac{V}{V_{\rm o}}\right)^{2}\right]^{0.5} } ~~ 
{\rm erf}(x_{\rm max}) ~~~{\rm mJy}
\end{equation}

\noindent where

$$ x_{\rm max} = \frac{2 \sqrt{ {\rm ln}2} R_{0}}{B} \left[1 -
  \left(\frac{V}{V_{\rm o}}\right)^{2}\right]^{0.5}$$

\noindent (see Knapp \& Bowers 1983; Olofsson et al. 1993). 
Here $f$ is the conversion factor between units of 
brightness temperature in kelvins
and mJy per beam for the radio telescope used, $d$ is the distance to the
star, $B$ is the half-power
synthesized beamwidth of the telescope, $V_{\rm o}$ is the outflow
velocity of the wind, $\mu$ is the mean atomic weight of the gas
(taken to be 1.3), and
$m_{H}$ is the mass of the hydrogen atom. All quantities are expressed
in cgs units.
 
\begin{figure}
\hspace{-0.6in}
\scalebox{0.7}{\rotatebox{0}{\includegraphics{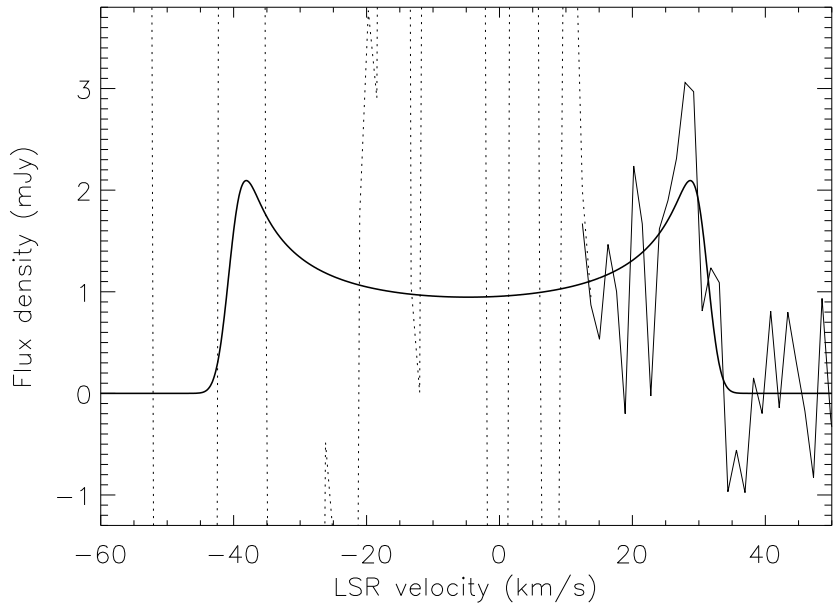}}}
\caption{Observed VLA \HI\ spectrum toward the position of \DC, 
averaged over
a single synthesized beam centered on the star (thin lines), with the
  best-fitting model \HI\ spectrum overplotted as a thick line (see
  \S~\ref{vlamassloss} for details). The observed spectrum was extracted
from the ``Robust+1'' data cube (see Table~3). 
The portion of the observed
  spectrum indicated by a dotted
  line is heavily contaminated
  by Galactic emission and was excluded from the fit. 
 }
\label{fig:modelspec}
\end{figure}

For comparison with the theoretical profile, we extracted from the VLA
data a spectrum, averaged over a single synthesized
beam centered on \DC. We used the
``Robust+1'' data cube for this analysis (Table~3), which has the
highest angular resolution and should
minimize contamination from structures in the \HI\
envelope that are expected to 
lie outside the freely expanding wind (e.g., material in the tail and/or
bow shock). We then compared the observed 
spectrum with a series of 
models where the mass-loss rate and wind outflow speed were taken as
free parameters. Before comparison with the data, 
each model computed from Eq.~1 was convolved with a broadening kernel
to account for the finite spectral resolution and the gas 
turbulence.  The  model parameters were constrained
by scanning a grid of possible values and minimizing $\chi^{2}$. 
The outflow velocity, mass-loss rate, and smoothing kernel were
incremented by values of 0.1~\kms,
$1\times10^{-8}~M_{\odot}$~yr$^{-1}$, and 0.1~\kms, respectively.
The best-fitting model profile is overplotted on the observed
spectrum in Figure~\ref{fig:modelspec}. The
model parameters are $V_{\rm o}=35.6\pm1.2$~\kms, ${\dot
  M}=(1.7\pm0.1)\times10^{-6}~M_{\odot}$~yr$^{-1}$, and $\sigma_{\rm
  turb}=1.8\pm0.2$~\kms. The quoted uncertainties do not reflect the
uncertainties in parameters held fixed in the model, namely the
stellar systemic velocity, the envelope radius, and the distance to
the star. Our model
underestimates the peak \HI\ flux density of the observed spectrum, 
which may reflect uncertainties in the fixed parameters of the model
(e.g., the envelope radius) or be due to
contamination of the observed spectrum by swept-up interstellar gas and/or
gas from outside the free expansion zone of the wind.
It may also reflect a breakdown of our highly idealized model (e.g.,
owing to highly episodic mass-loss or a latitude dependent wind).

A second, independent means of constraining the mass-loss rate from
the \HI\ data comes from the total mass of circumstellar \HI\
emission observed, coupled with available constraints on the timescale for the
mass-loss.  Over the velocity range $13.8\le V_{\rm LSR}\le
38.2$~\kms, 
we directly measured a total \HI\ mass of 0.07~$M_{\odot}$ from
material that appears to be associated with the
circumstellar environment of \DC\ (see \S~\ref{results}). 
Applying a multiplicative correction of 1.34 to account for the mass of
helium yields $M_{\rm tot}\gsim 0.09~M_{\odot}$. As noted above, 
we regard this as a
lower limit to the total envelope mass, since additional circumstellar
emission may lie in portions of our band that are contaminated by
Galactic signal and/or at large angular scales to which the VLA is insensitive.

The tail of \HI\ emission downstream from \DC\ has an extent of
$\sim500''$, or $2\times10^{18}$~cm in the plane of the sky. Assuming
a tangential velocity of 7~\kms\ (derived from the observed stellar
radial velocity and proper motions as described in Marengo et
al. 2010b and the new solar constants of Sch\"onrich et al. 2010), 
the crossing time for the
nebula is $\sim$90,600~yr. For the total circumstellar mass above, 
this would imply a mean
mass-loss rate of ${\dot M}\approx 1\times
10^{-6}~M_{\odot}$~yr$^{-1}$. While this is comparable to the rate derived
from our model fitting above, the actual mass-loss rate
could be smaller if some fraction of the observed \HI\ gas
arises from swept-up interstellar material rather than directly from
stellar mass-loss (see \S~\ref{sweeping} below).
Moreover, the dynamical crossing time of a stellar wake
provides only a lower limit to its age (see Matthews et al. 2008); if
\DC\ has been losing mass over a more prolonged interval, ${\dot M}$
would also be lower. 
We note, however, that
\DC\ is believed to be on its second
crossing of the Cepheid instability strip (Berdnikov et al. 2000;
Turner et al. 2006), and the total
duration of this phase for a star of its mass and metallicity 
is expected to be $\sim$(1.3-5.9)$\times10^{5}$~yr (see \S~\ref{resolution}). 
These numbers are only a few times larger than our estimated dynamical age,
suggesting that the mean mass loss rate cannot be significantly lower
than we have inferred
under the assumption that all of the observed mass-loss occurred during the
Cepheid phase. [The duration of the first Cepheid 
crossing ($\sim15,000-30,000$~yr for a star like 
\DC; see \S~\ref{resolution}) is too short to
impact this conclusion].
On the other hand, if we have underestimated the amount of
emission in the \HI\ nebula because of Galactic confusion, this would
imply a {\em higher} mass-loss rate, although this correction is
expected to be less than a factor of $\sim$2 or 3, assuming the line
profile is symmetric. 

\subsubsection{Is a Significant Fraction of the Circumstellar Material 
Swept from the ISM?\protect\label{sweeping}}
As mass-losing stars move through the ISM, their jetsam
can be augmented by material swept
from the local ISM. The fraction of the circumstellar envelope
that originates in this manner will depend on a combination of the
stellar mass, the stellar wind
parameters, and the ambient ISM density 
(Villaver et al. 2002). Accounting for this swept-up interstellar
mass is important in evaluating the true mass-loss rate from the
star.

In the absence of a detailed numerical model, we can obtain a crude estimate
of the amount of swept-up material in the circumstellar
environment of \DC\ as 
$M_{s}=\int^{t_{ml}}_{0}\pi r^{2}V_{s}\rho~dt$,
where $r$ is the effective radius of the star, $V_{s}$ is its
space velocity, $\rho$ is the mass density of the ambient medium, 
and $t_{ml}$ is the length of time that the star has been
losing mass. 

Adopting an ISM particle density $n\approx$10~cm$^{-3}$
(see \S~\ref{pastcomp}) translates
to $\rho=1.7\times10^{-23}$~g~cm$^{-3}$.
Assuming $r$=0.107~pc (Marengo et al. 2010b), $V_{s}$=10.3~\kms\
(\S~\ref{introduction}), and $t_{ml}$=90,600~yr (\S~\ref{rates}), this yields
$M_{s}\approx0.01(n_{\rm HI}
/10~{\rm cm}^{-3})(t_{ml}/90,600~{\rm yr})~M_{\odot}$, or roughly 14\% of the
directly observed \HI\
mass.  Even after allowing for 
factors of a few uncertainty in both $n$ and
$t_{ml}$, this supports our interpretation of the emission detected by
the VLA as arising predominantly from a wind from \DC. This is also
consistent the conclusions of Marengo et al. (2010b), who argued that
the low inferred PAH content of the material detected by
{\it Spitzer} implies that it is dominated by 
debris originating in the stellar wind. In the future, more sophisticated
numerical simulations  (cf. Villaver et al. 2002) could help to
obtain more accurate constraints on $M_{s}$. 

\subsubsection{Comparison with Previously Derived 
Mass-Loss Rates for \DC\protect\label{pastcomp}}
Based on their {\it Spitzer} measurements, 
Marengo et al. (2010b) used two independent approaches to
estimate a mass-loss rate for \DC\ in the range
${\dot M}\approx5\times10^{-9}$ to
$6\times10^{-8}~M_{\odot}$~yr$^{-1}$.  If we scale these values to
account for revised stellar space velocity of 
$V_{s}$=10.3~\kms\ adopted in the present
work (see \S~\ref{introduction}) and
the outflow velocity of $V_{\rm o}\approx$35.6~\kms\ inferred from the \HI\
data, this translates to ${\dot M}\approx7.2\times10^{-9}$ and
$2.1\times10^{-8}~M_{\odot}$~yr$^{-1}$, respectively. 
These values are consistent with theoretically predicted mass-loss
rates for Cepheids (Neilson \& Lester 2008), but are
significantly lower than
the values that we infer from the \HI\ data.

The smaller of the two mass-loss rate values from Marengo et
al. (2010b) was derived based on
the observed stand-off distance of the bow  shock structure together
with ram pressure balance arguments. The implied mass-loss rate could
be higher if the local ISM particle density in the vicinity of \DC\ is larger
than the originally adopted value of $n=$0.55
cm$^{-3}$. Indeed, there is evidence that a higher particle density in
the region
is plausible. Fernie (1990)
derived a reddening value of $E(B-V)=0.092$ toward \DC; assuming
$A_{V}/E(B-V)$=3.1 (Savage \& Mathis 1979), 
this implies a $V$-band extinction
coefficient $A_{V}\approx0.29$.  Such a value is typical of
interstellar clouds with $n\approx$10-50~cm$^{-3}$  
(Turner 1997). We note that owing to its larger
angular extent, the circumstellar 
dust detected by {\it Spitzer} should contribute
negligibly to $A_{V}$. Assuming a typical grain radius $a$=0.001$\mu$m, a
total dust mass $M_{\rm dust}$=$6\times10^{-7}~M_{\odot}$, and a 
radial extent $r$=0.103~pc (Marengo et al. 2010b), 
then $A_{V,{\rm local}}\sim0.22 M_{\rm dust}a^{-1}r^{-2}\approx 0.01$.  
Finally, from the database of 
Kalberla et al. (2005), the \HI\ column density
along the line-of-sight to \DC\ is $N_{\rm
  HI}=7.7\times10^{21}$~cm$^{-2}$, implying a mean particle density
out to the distance of the star of 
$n\approx$9.2 cm$^{-3}$. This is again
consistent with a local  particle density several times higher than
the canonical Galactic plane value for warm, intercloud \HI\ 
(cf. Dickey \& Lockman 1990). Assuming a local ISM density with
$n\sim10$~cm$^{-3}$ would imply ${\dot M}\sim
1.4\times10^{-7}~M_{\odot}$~yr$^{-1}$ based on ram pressure balance
arguments, in somewhat better agreement with our \HI-derived value given
the uncertainties in both estimates. 

The second mass-loss rate estimate from Marengo et al. (2010b) 
was derived under the assumption that the 
gas-to-dust ($g/d$) ratio of the circumstellar material 
has a value of 100, typical of what is found
 in the ISM and
in the circumstellar environments of AGB stars.
However, it is presently unknown 
whether this value should hold in the circumstellar
environment of a Cepheid. 
For example, the ratio $g/d$ is found
to be 5-10 times higher than the canonical value 
in the circumstellar envelopes of some
supergiants (e.g., Skinner \& Whitmore 1988),
presumably because grain condensation has been incomplete 
in the winds of such stars (Mauron \& Josselin 2011).  Even more
extreme values may be plausible in the winds of Cepheids owing to
their higher temperatures and higher ultraviolet fluxes. In the case of
\DC, $g/d\approx 2300$ would be needed to
reconcile the mass-loss rate of Marengo et al. (2010b) with our \HI\
estimates.
Interestingly, such a value would be self-consistent in the sense
that the empirically derived 
mass volume density of the dusty wind, $\rho_{\rm OF}$,
would then match the density
for a wind with ${\dot M}\approx1\times10^{-6}~M_{\odot}$~yr$^{-1}$
and $V_{\rm o}$=35.6~\kms\ as predicted by the mass continuity equation for
stellar winds: ${\dot M}=4\pi\rho_{\rm OF}r^{2}V_{\rm o}$ (Lamers \&
Cassinelli 1999).

In summary, our best estimate of the mass-loss rate of
\DC\ based on the combined results from our VLA \HI\ data and {\it
  Spitzer} infrared data is ${\dot
  M}\approx(1\pm0.8)\times10^{-6}~M_{\odot}$ yr$^{-1}$. 
The quoted error bar reflects the dominant systematic uncertainties
and is based on the dispersion in the values derived
from the independent methods presented in the preceding sections.

\subsection{Implications of the Empirically Derived Mass-Loss Rate
  for Theoretical Models of Cepheids\protect\label{neilson}}
\subsubsection{Consistency with Theoretically Predicted Mass Loss
  Rates}
The majority of Cepheids for which mass-loss rates (or upper limits) 
have previously
been determined (\S~\ref{introduction}) 
are inferred to have ${\dot M}$ values at least two or three 
orders of magnitude smaller than we
have derived for \DC\ in the present work. 
We emphasize however, that none of these previous
determinations exclude the possibility of significant mass-loss
through a predominantly neutral atomic wind, of the type we have
detected
through \HI\ observations. In the case of \DC, the wind also
revealed its presence through FIR emission, although this need not be
the case for all mass-losing Cepheids, particularly those with higher effective
temperatures (where dust grains might be destroyed more readily),
those that lack a companion (which may play a role in exciting FIR
emission), or
those residing in lower density environments (where dust swept from the
ISM might contribute negligibly to the circumstellar FIR emission). It
is clear that an \HI\ survey of a larger sample of Cepheids will be
needed to obtain a better understanding of the importance of 
these factors and to assess the ubiquity of Cepheid mass loss that is
traceable through \HI\ 21-cm line emission.

Based on theoretical studies to date, the most promising 
physical mechanism by which Cepheids could lose significant 
mass appears to be the
enhancement of radiatively driven mass loss through shocks and 
pulsation 
(Willson \& Bowen 1984; Neilson \& Lester 2008). 
From analytic models of this type, Neilson \& Lester (2008) and
Neilson et al. (2011) predicted
typical Cepheid
mass-loss rates in the range 10$^{-10}$ to
10$^{-7}~M_{\odot}$~yr$^{-1}$, although these rates are highly
sensitive to the balance of forces and the adopted
physical parameters for a given star. In particular, for shorter
periods (smaller masses),  the mass-loss rate has a
non-linear dependence on the stellar mass. Thus for a Cepheid with the
mass and period of \DC, it appears that 
the models of Neilson \& Lester and Neilson et al. can
in principle accommodate
mass-loss rates as high as $10^{-6}~M_{\odot}$~yr$^{-1}$, consistent
with our VLA findings (H. Neilson, private communication). However, this
conclusion may change if mass loss is highly episodic (see below).

\subsubsection{The Effects of Mass Loss on the Pulsation Period}
Secular period changes are well known to occur in Cepheids,
but are generally assumed 
to be dominated by evolutionary effects (which result in
changes in the effective temperature and luminosity) rather than mass loss 
(Turner et al. 2006). 
It is therefore of considerable interest to compare the
change in pulsation period for \DC, ${\dot P}$, that is predicted to occur as
a result of its derived
mass loss rate with the value of ${\dot P}$ previously 
derived from observational data.

Equation~62 of Neilson \& Lester (2008) provides an analytic formula for
estimating the effect of mass loss on the rate of change
in the stellar pulsation period. The most
extreme period changes due to mass loss are expected to occur when 
changes in both $T_{eff}$
and $L$ (and hence $R$) 
are minimal. Moreover, the time scale for the decrease in
total mass is predicted to be faster than the canonical evolutionary time
scale. Under these conditions, the period changes 
dominated by mass-loss should always be positive. If we take the
pulsation period and pulsation mass for \DC\ from Table~1, together with ${\dot
  M}=10^{-6}~M_{\odot}$~yr$^{-1}$, using Equation~62 of Neilson \&
Lester we estimate the
period change as a result of mass loss from \DC\ to be ${\dot
  P}=0.06$~s~yr$^{-1}$. This is in apparent contradiction to the
observed value
of ${\dot P}=-0.1$~s~yr$^{-1}$ derived for \DC\ by Berdnikov et
al. (2000) based on data spanning 130 years. This suggests that mass
loss is only one contributor to the secular period change and that
evolutionary effects on the star following mass loss 
cannot be ignored.
An additional consideration is that mass loss inside the instability
strip may be episodic. This could explain, for example, why most previous
searches for Cepheid mass loss (which have only been sensitive to
material in close proximity of the star) have typically implied much
lower rates of mass-loss than we infer from the VLA and {\it Spitzer}
observations of \DC. 
However, in order to account for the same total
amount of matter shed, highly episodic mass loss
would require a boost in efficiency of the Cepheid mass loss mechanism.

\subsubsection{Could the Mass Loss Have Occurred During an Earlier
  Evolutionary Phase?}
One of the few Cepheids that has been previously inferred to have a 
mass loss rate comparable to
the rate we have derived for \DC\  in our present study is RS~Pup, where ${\dot
  M}\approx 10^{-6}~M_{\odot}$~yr$^{-1}$ (Deasy 1988; Barmby et
al. 2011). RS~Pup
is unique among Galactic Cepheids in that it is surrounded by a
prominent reflection nebula. The linear extent of this nebula 
($\sim$1~pc) is comparable to the \HI\
nebula that we measure toward \DC.  Although the origin of RS~Pup's nebulosity
is consistent with copious 
mass loss, some authors have argued that
the bulk of the nebular material may have originated during an earlier
evolutionary phase, either as a Be star (Kervella et al. 2009) or as a red
giant (Havlen 1972). The question of whether such arguments could also
be applied to \DC\ is important,
since it impacts whether subsequent mass loss during the Cepheid phase 
will be sufficient
to reconcile the Cepheid mass discrepancy for this star. 

The geometry of the nebula that we observe surrounding \DC\ places
strong constraints on this issue. The presence of an
infrared-emitting bow shock and a head-tail structure to the \HI\
nebula strongly suggest that \DC\ has {\em ongoing}
mass-loss. While this alone does not exclude the
possibility that some portion of the material in its circumstellar nebula was
shed during an earlier evolutionary phase (e.g., at the tip of
the red giant branch), the fact that the dynamical
age of the \HI\ nebula is smaller than the predicted duration of the second
crossing of the instability strip for \DC\ (see \S~\ref{resolution}) suggests a
strong likelihood that the bulk of the mass loss occurred during the
Cepheid evolutionary phase. Further, the dynamical age of the nebula
implies that even if some of the 
mass were lost during the red giant phase, this
is likely to have immediately preceded the current 
Cepheid crossing. Hence the resulting mass loss
would still necessitate a change in the evolutionary mass of the star
between its first and its second (current) Cepheid crossing.

\subsection{Comments on the Inferred Outflow Velocity for the \DC\
  Wind and Its Implication for the Mass Loss Mechanism\protect\label{Vout}}
In \S~\ref{vlamassloss}, we estimated an asymptotic (outflow) 
velocity for the \DC\ wind of
$V_{\rm o}\approx$35.6~\kms. To our knowledge, this is the first
directly measured outflow velocity for a Cepheid. 
An important characteristic of this value is
that it is significantly smaller than the stellar escape velocity.
Adopting the mean stellar radius and pulsation mass for \DC\ quoted in
Table~1, 
the escape velocity from the photospheric surface of \DC\ is $V_{\rm
 esc}\approx 200$~\kms, or several times the asymptotic velocity. 
While the radius and effective surface gravity of \DC\ change over the
course of its pulsation cycle, these effects are not sufficient to change this
conclusion.  Andrievsky et al. (2005) reported
changes in the effective surface gravity of \DC\ by a factor of 5.6,
which implies that even when the surface gravity is at its minimum,
the escape velocity would still be $V_{\rm esc}\approx120$~\kms. \DC\ 
thus appears to share one of the defining
characteristics of the cool winds of
AGB stars and late-type supergiants, where measured outflow velocities
are in general several times smaller than the photospheric escape
speed (i.e., $V_{\rm o}<<V_{\rm esc}$; 
see Holzer \& MacGregor 1985; Judge \& Stencel 1991). Interestingly,
\DC\ also appears to adhere to the empirical relation
$V_{\rm o}\sim1.6\times10^{-3}V_{\rm esc}^{2}$  found by Reimers (1977)
based on observations of K giants and G and K supergiants (see also
Judge 1992), although the 
significance of this is unclear, since
the physical underpinnings of this relation are poorly understood
(Reimers 1977) and \DC\ is not expected to be undergoing mass loss via
the same mechanism as the stars in Reimers' sample. 

As discussed by Holzer \& MacGregor (1985), the condition that 
$V_{\rm o}<<V_{\rm esc}$ has important implications for understanding and
constraining 
the mechanism(s) driving a stellar wind. For example, this
requires the existence of some type of regulatory processes such 
that most of the driving energy of the wind
goes into lifting material out of the stellar gravitational field
rather than accelerating the flow. It
follows that the inferred momentum
flux of the wind, ${\dot M}V_{\rm o}$, represents only a small fraction
of the overall energy required to drive the wind. An additional
implication is that most of the energy added to the
wind must be in the form of momentum rather than heat.

\subsection{A Possible Solution to the Cepheid Mass Discrepancy 
for \DC?\protect\label{resolution}}
A fundamental question raised by our new empirically derived mass-loss
rate for \DC\ is whether this degree of mass loss can
reconcile the mass discrepancy for the star. As described above,
\DC\ is believed to be on its second
crossing of the instability strip. The total
duration of this evolutionary 
stage (as well as for the first and third crossings) has previously 
been calculated for Cepheids
with different masses and metallicities by Bono et al. (2000a). 
However, these models did not include the effects of mass loss and
therefore their direct application of the case of \DC\ will not yield
self-consistent results. To overcome this problem, we computed new intermediate-mass
evolutionary
models specifically for $\delta$~Cephei. The theoretical framework we
employed was developed by Pietrinferni et al. (2004). The
evolutionary
tracks we adopted are available in the BaSTI data base\footnote{A few
selected
evolutionary tracks were specifically computed by A. Pietrinferni for this
project.} and
were constructed by assuming a scaled solar
chemical mixture
and a fixed solar chemical composition  with helium and metal abundances of
$Y$=0.273 and $Z$=0.02, respectively. 

Several lines of empirical evidence support
the occurrence either of mild convective core overshooting during the
central hydrogen
burning phases of intermediate mass stars (Barmina et al. 2002; Cordier et
al. 2002;
Cassis \& Salaris 2011; Evans et al. 2011) or of enhanced post-main
sequence mass loss
(Bono et al. 2002; Natale et al. 2008; Cantiello et al. 2011).
However, to
constrain the effects of different mass loss rates, our current canonical
evolutionary
models were constructed by neglecting extra-mixing sources both at the
edge of the
convective core and at the bottom of the convective envelope (Alongi et
al. 1991).
We plan to provide a more detailed discussion of the impact of the mass
loss on
the evolutionary properties of helium burning intermediate-mass stars in a
forthcoming paper (Bono et al., in preparation).

\begin{figure}
\hspace{1.0in}
\scalebox{0.7}{\rotatebox{0}{\includegraphics{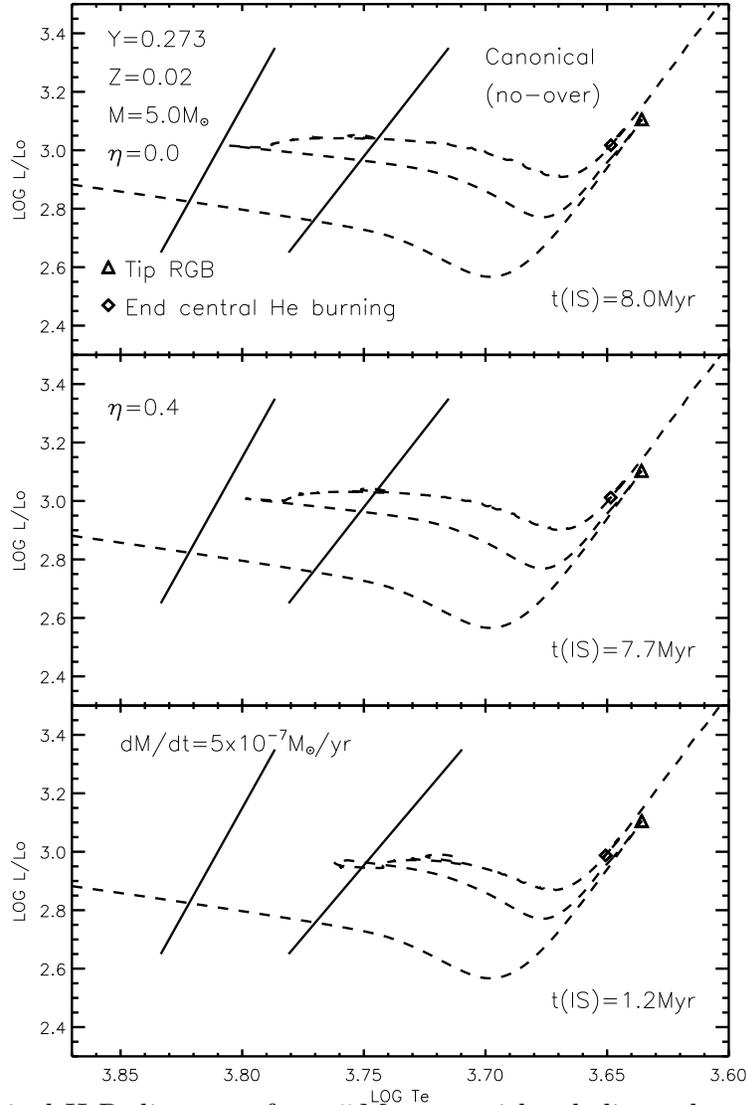}}}
\caption{Theoretical H-R diagrams for a 5$M_{\odot}$ star with a helium
 abundance $Y$=0.273 and metal abundance $Z$=0.02. The dashed lines
 indicate the evolutionary path of the star. The solid lines depict
 the boundaries of the instability strip. The top panel shows a
 canonical model with no mass loss. The center panel includes mass
 loss starting on the main sequence and following 
a Reimers law with $\eta=$0.4 (Eq.~2). The lower panel shows a model where
 mass loss with a rate ${\dot M}=5\times10^{-7}~M_{\odot}$ yr$^{-1}$
 turns on at the red edge of the instability strip and continues
 throughout the Cepheid phase. The total time spent on the instability
 strip for each of the models is indicated in the lower right corner. 
No overshoot is included. See \S~\ref{resolution} for
 further details. }
\label{fig:tracksa}
\end{figure}

Figure~\ref{fig:tracksa} shows H-R diagrams for evolutionary tracks
at a fixed initial stellar mass ($M_{\star}=5M_{\odot}$)
but different assumptions concerning the mass-loss
rate. Each panel shows the
evolutionary phases just before and soon after helium burning,
i.e. the so-called ``blue loop''. The solid lines denote the predicted
first overtone blue (hot) edge and the fundamental red (cool) edge of the
Cepheid
instability strip according to Bono et al. (2000b). The triangle and the
diamond
mark the onset (tip of the red giant branch) and the end of central
helium burning phases, respectively. The evolutionary time spent inside the
instability strip is also indicated. The evolutionary track plotted in the top
panel was constructed by neglecting mass loss. The
track plotted in the middle panel was constructed by assuming a mass loss
starting from the main sequence phase according to the semi-empirical formula
from Reimers (1975):

\begin{equation}
{\dot M} = 4\times10^{-13} \eta
\frac{(L/L_{\odot})(R/R_{\odot})}{M/M_{\odot}}
\end{equation}

\noindent The free parameter $\eta$ was fixed at 0.4 (Pietrinferni et al.
2004;
Bono et al. 2006). Following this prescription, the mass loss rate during the
evolution of the star will vary according to the changes in
surface luminosity and radius. However, a comparison of the 
theoretical predictions plotted in the
two top panels of  Figure~\ref{fig:tracksa}
shows that the impact of the mass loss 
based on the prescription 
of Reimers is negligible, and indeed, both the extent in effective
temperature of the blue loop and the evolutionary time spent inside the
instability strip are minimally affected. Plausible increases in the
free parameter $\eta$ do not change this outcome, and the Reimers
model cannot produce sufficient mass loss to account for the observed
quantity of circumstellar material around \DC, nor can it reconcile the mass
discrepancy for this star.

Finally, in the bottom panel of Figure~\ref{fig:tracksa},
we show an evolutionary track with the same initial
mass of the top panels, but assuming a steady mass loss rate of
${\dot M}=5\times10^{-7}~M_{\odot}$~yr$^{-1}$
during the evolutionary phases inside the Cepheid instability strip.
As expected, we found that implementing such mass-loss causes a
significant
decrease in the extent in effective temperature of the blue loop owing
to the decrease in the envelope mass (Bono et
al. 2000a,
and references therein), and in turn a decrease in the evolutionary time spent inside
the instability 
strip. We also explored models with higher mass-loss rates (not shown), but 
found that
for ${\dot M}\gsim10^{-6}~M_{\odot}$~yr$^{-1}$, the temperature extent
of the blue loop becomes
significantly smaller,  leading to implausibly short Cepheid lifetimes.

\begin{figure}
\hspace{1.0in}
\scalebox{0.7}{\rotatebox{0}{\includegraphics{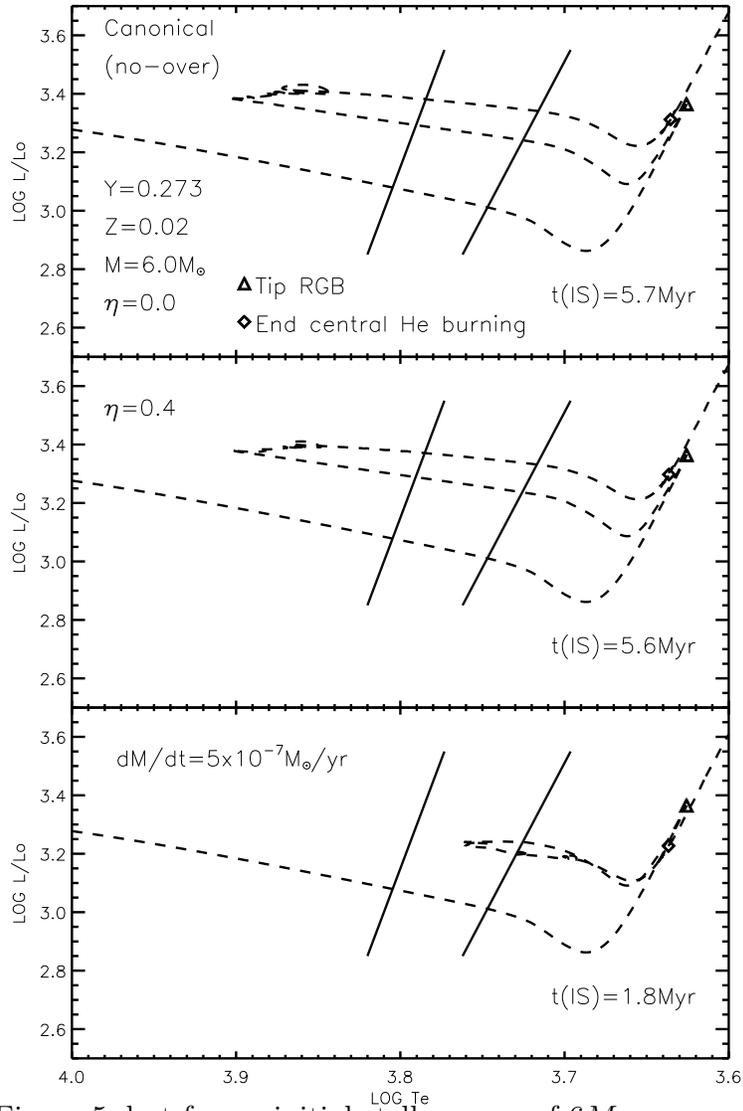}}}
\caption{As for Figure~\ref{fig:tracksa}, but for an initial
 stellar mass of 6$M_{\odot}$.
 }
\label{fig:tracksb}
\end{figure}

To further constrain the impact of the initial stellar mass, we
computed an additional
set of evolutionary models with an initial stellar mass 
$M_{\star}=6M_{\odot}$ (Figure~\ref{fig:tracksb}). The evolutionary tracks
show a behavior similar to the $M_{\star}=5M_{\odot}$ models when
moving
from the case of 
no mass-loss (top) to the Reimers mass-loss case (middle) to
the case of mass loss inside the
instability strip at a rate of ${\dot
  M}=5\times10^{-7}~M_{\odot}$~yr$^{-1}$ (bottom). As with the
$M_{\star}=5M_{\odot}$ models, higher mass-loss rates 
(${\dot M}\gsim10^{-6}~M_{\odot}$~yr$^{-1}$) destroyed the blue loop,
resulting in a negligible lifetime on the instability strip.

The above findings indicate that either (1) the true mass-loss rate from
$\delta$ Cephei
is at the low end of our empirically derived range,
or (2) that
a more complex treatment of mass loss is required
(e.g., episodic mass loss or implementation of multi-dimensional 
Eulerian hydrodynamic and stellar evolution codes; Moc\'ak et
al. 2009).
The inclusion of extra mixing and stellar rotation may also be important.
We consider possibility (1) below, while possibility (2)
will be explored in future works.

If we assume that \DC\ is currently near the midst of its second crossing
of the instability strip, the models presented in Figures~\ref{fig:tracksa} and
\ref{fig:tracksb} predict that a Cepheid with the chemistry and evolutionary
mass of \DC\ and undergoing mass loss at a rate of ${\dot
  M}=5\times10^{-7}~M_{\odot}$~yr$^{-1}$ will have so far spent roughly
(0.9-3.0)$\times10^{5}$~yr on the instability strip (most of it on the
second crossing). These numbers are consistent with the dynamical age
of the \HI\ wake estimated in \S~\ref{vlamassloss}.
Steady mass-loss at the prescribed rate over such an interval is
predicted to result in a mass of circumstellar material of $\sim$0.04
to 0.15$M_{\odot}$. 
The upper end of this range is consistent with our empirically derived
estimate for the mass of circumstellar material after correction for
He (\S~\ref{vlamassloss}),
implying that such a model is self-consistent. 
Moreover, this amount of material is
remarkably close to the change in mass required to reconcile the
pulsation and evolutionary masses of \DC\ 
($M_{e}-M_{p}\approx 0.2-1.8~M_{\odot}$ assuming that 
$M_{e}=5.5\pm0.5~M_{\odot}$. We conclude that
to within current observational and theoretical uncertainties,
{\em mass loss offers a tenable solution to resolving a significant
  fraction of the 
the mass discrepancy for \DC.} 

From observations of only a single star, we cannot
draw any general conclusion about the ubiquity or evolutionary
importance of mass loss from
Cepheids. Furthermore,
as discussed in \S~\ref{introduction}, a general solution to the
mass discrepancy for all Cepheids is likely to require a combination of
both mass-loss and overshoot. However, since only
the mass loss can be observed directly, future efforts to identify
large-scale  
circumstellar debris around a larger sample of 
Cepheids and a comparison of these results with theoretical 
models
(including evolutionary models and non-linear convective hydrodynamic models)
are likely to provide powerful new constraints on the
relative contributions of mass loss versus overshoot 
for Cepheids of various masses.

\section{Conclusions}
We have used the VLA to search for \HI\ 21-cm line emission in the
circumstellar environment of the archetype of Cepheid variables,
\DC. We have detected an extended ($13'$, or $\sim$1~pc) nebula at the
position of the star.  The nebula exhibits a head-tail morphology, with the
head of the structure aligning closely with the infrared-emitting 
nebula and bow shock previously
detected by Marengo et al. (2010b), while the tail appears
to be a turbulent structure that 
trails the motion of the star through the ISM. We measure 
an \HI\
mass for the nebula of $M_{\rm HI}\approx0.07~M_{\odot}$, although its
total \HI\ mass could be
$\sim$2-3 times larger, depending on the fraction of emission that is
hidden by the strong Galactic emission that contaminates a portion of
our observing band. Additional material may
also be present on large angular scales ($\gsim15'$) to which the VLA
is insensitive. We interpret the bulk of the
\HI\ nebula
surrounding \DC\ as arising from a stellar wind with a mass-loss rate
of ${\dot M}\approx(1.0\pm0.8)\times10^{-6}~M_{\odot}$~yr$^{-1}$ 
and an outflow velocity
$V_{\rm o}\approx35$~\kms. 
By computing evolutionary models that include mass loss across the
instability strip, we show that
a mass-loss rate of this magnitude, sustained over
the preceding Cepheid lifetime of \DC, could be sufficient to resolve 
a significant fraction of the
longstanding discrepancy between the masses of the star derived from
stellar 
pulsation versus stellar evolutionary models. 

\acknowledgements
It is a pleasure to thank A. Pietrinferni for the computation of several
evolutionary tracks and for many insightful discussions concerning the
evolutionary properties of intermediate-mass stars. 
We are also grateful to H. Neilson for valuable discussions concerning Cepheid mass loss.
The observations presented here were obtained through NRAO program AM998.
LDM acknowledges support for this work from 
grant AST-1009644 from the National Science Foundation. NRE
acknowledges support from the Chandra X-ray Center, NASA
contract NAS8-03060.

\clearpage

\end{document}